# Biochemical Oscillations in Delayed Negative Cyclic Feedback: Harmonic Balance Analysis with Applications [*]

Yutaka Hori    Shinji Hara [†]

**Abstract**

Oscillatory chemical reactions often serve as a timing clock of cellular processes in living cells. The temporal dynamics of protein concentration levels is thus of great interest in biology. Here we propose a theoretical framework to analyze the frequency, phase and amplitude of oscillatory protein concentrations in gene regulatory networks with negative cyclic feedback. We first formulate the analysis framework of oscillation profiles based on multivariable harmonic balance. With this framework, the frequency, phase and amplitude are obtained analytically in terms of kinetic constants of the reactions despite the nonlinearity of the dynamics. These results are demonstrated with the Pentilator and Hes7 self-repression network, and it is shown that the developed analysis method indeed predicts the profiles of the oscillations. A distinctive feature of the presented result is that the waveform of oscillations is analytically obtained for a broad class of biochemical systems. Thus, it is easy to see how the waveform is determined from the system's parameters and structures. We present general biological insights that are applicable for any gene regulatory networks with negative cyclic feedback.

## 1  Introduction

In living cells, oscillatory chemical reactions serve as a timing clock of important cellular processes. The temporal dynamics of oscillatory gene expression has thus been actively studied in recent years (Goldbeter and Berridge, 1997; Winfree, 2001). It is known that the frequency, phase and amplitude of the oscillations are diverse, ranging from minutes to hours and from phase synchronization to asynchronous oscillations. However, the relation between the dynamical properties of biochemical system and the resulting temporal pattern is not thoroughly understood. Here we propose a theoretical framework for quantitatively studying the frequency, phase and amplitude of oscillatory chemical concentrations that arise from biochemical networks with negative cyclic feedback.

The negative cyclic feedback motif shown in Fig. 1, where each gene activates or represses another gene expression in a cyclic way, has been considered as a core circuit module to produce periodic oscillations for a long time (Goodwin (1965); Tyson (1975); Thron (1991); Tiana et al. (2007); Hori et al. (2011) for example). This conjecture was recently corroborated with a synthetic biological circuit reported in Elowitz and Leibler (2000), in which oscillatory gene expression was indeed observed in a gene regulatory network consisting of three repressors. Moreover, the negative cyclic feedback was also found in many existing biochemical networks that exhibit periodic oscillations such as the somitogenesis oscillator (Hirata et al., 2004) and p53 networks (Lahav et al., 2004). Thus, the study of negative cyclic feedback circuits can potentially unravel the essential dynamical properties of oscillatory biochemical reactions.

---

[*]Corresponding author Y. Hori. Tel. +81-3-5841-6893. Fax +81-3-5841-7961.

[†]Y. Hori and S. Hara are with Department of Information Physics and Computing, The University of Tokyo, 7-3-1 Hongo, Bunkyo-ku, Tokyo 113-8656, Japan. {Yutaka_Hori, Shinji_Hara}@ipc.i.u-tokyo.ac.jp

The frequency profile of oscillatory gene expression in cyclic biochemical networks was studied by means of numerical simulations for the Repressilator motif (El-Samad et al., 2005) and for the self-negative feedback motif (Jensen et al., 2003; Monk, 2003; Wang et al., 2004). A more theoretical approach was taken by Rapp (1976) to investigate the frequency of Goodwin type oscillators (Goodwin, 1965) based on harmonic balance analysis (Khalil, 2001). This approach is, however, not directly applicable to the biochemical oscillators of our interest due to the multiple nonlinearities, and the phase profile of oscillations is still open to be analyzed. Hence, a more general theoretical framework is desirable to systematically explore the profiles of oscillations in biochemical networks.

The goal of this paper is twofold: (i) to develop a systematic method to study the profiles of oscillations in gene regulatory networks with negative cyclic feedback, and (ii) to gain novel biological insight on the relation between the oscillation profiles and the reaction kinetics. In particular, we here pursue universal insights that are applicable to a broad class of biochemical networks rather than the numerical study of particular biological examples. To this end, we adopt a simplified model that is suitable for capturing the essence, and we analytically obtain the oscillation profiles. Specifically, we use describing function approach to approximating nonlinearities of the system (Khalil, 2001; Iwasaki, 2008), and formulate multivariable harmonic balance equations for the analysis of biochemical oscillators. Using distinctive features of the dynamics of biochemical networks, we derive analytic estimates of frequency, phase and amplitude in terms of kinetic constants of the reactions despite the nonlinearity of the dynamical model.

A distinctive feature of our result is that the approximate waveform of oscillations is analytically obtained for a broad class of cyclic biochemical systems. Thus, it is easy to see how the waveform is determined from the system's parameters and structures. This feature is demonstrated through the analysis of two biological examples, the Pentilator (Tsai et al., 2008) and the somitogenesis clock (Hirata et al., 2004). The Pentilator is a conceptual biochemical oscillator with five repressors connected in a cyclic way. We here analyze the Pentilator to show that the developed result is useful even for large-scale gene regulatory networks. In the somitogenesis example, we investigate the Hes7 self-repression network (Hirata et al., 2004). It was experimentally observed that Hes7 gene exhibits oscillatory expression with almost two-hour cycle (Hirata et al., 2004). We demonstrate that the developed analysis method indeed predicts the two-hour period, and show that transcription and translation time delays play an important role in maintaining the oscillation period. Finally, general biological insights that are applicable to any biochemical networks with negative cyclic feedback are presented.

This paper is organized as follows. In Section 2, the dynamical model of the gene regulatory network is introduced. Then, we formulate multivariable harmonic balance for biochemical oscillators in Section 3. Based on this formulation, the frequency, phase and amplitude of oscillations are analytically obtained in Section 4. Section 5 is devoted to illustrate these results with the biological examples. In Section 6, we provide general biological insights. Finally, concluding remarks are given in Section 7.

The basic idea of the analysis was previously demonstrated in the authors' conference paper (Hori and Hara, 2011) with the omission of rigorous proofs. In this paper, we include complete proofs of the theorems. Moreover, the analysis of oscillation amplitude (Section 4.3) and the illustrative examples of the two biochemical oscillators (Section 5) are presented as original work. The biological insights in Section 6 are also greatly extended.

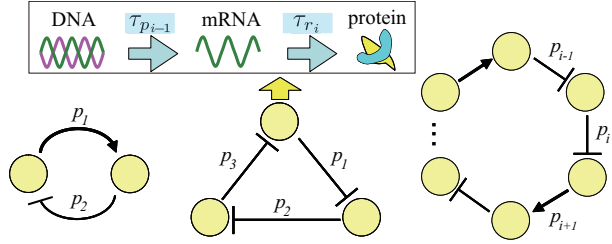

Figure 1: Gene regulatory networks with negative cyclic feedback. The symbols → and ⊣ represent activation and repression of transcription, respectively. (Left) activator-repressor motif, (Center) Repressilator motif (Elowitz and Leibler, 2000), (Right) generic negative cyclic motif considered in this paper.

## 2 Model Description and Existence of Oscillations

### 2.1 Dynamical Model of gene regulatory networks with negative cyclic feedback

In cells, the production of proteins consists of two processes, transcription and translation. In transcription, information on a gene is decoded, and turns into the copies of messenger RNA (mRNA). The mRNA molecules are then translated to corresponding protein molecules (see Fig. 1). These proteins then activate or repress the transcription of other genes, and form gene regulatory network.

In this paper, we consider the gene regulatory network illustrated in Fig. 1. Here each protein activates or represses another transcription in a cyclic way. The dynamics of mRNA and protein concentrations in such networks can be modeled as follows (Chen and Aihara, 2002; Elowitz and Leibler, 2000).

$$\begin{aligned}
\dot{r}_i(t) &= -a_i r_i(t) + \beta_i f_i(p_{i-1}(t - \tau_{p_{i-1}})), \\
\dot{p}_i(t) &= c_i r_i(t - \tau_{r_i}) - b_i p_i(t),
\end{aligned} \qquad (1)$$

for $i = 1, 2, \cdots, N$, where $r_i \in \mathbb{R}_+ (:= \{x \in \mathbb{R} \mid x \geq 0\})$ and $p_i \in \mathbb{R}_+$ denote the concentrations of the $i$-th mRNA and its corresponding protein synthesized by the $i$-th gene, respectively. For the sake of notational simplicity, we regard the subscript 0 as $N$ throughout this paper. The positive constants $\tau_{r_i}$ and $\tau_{p_i}$ describe transcription and translation time delays resulting from unmodeled intermediate process, respectively. The kinetic constants $a_i, b_i, c_i$ and $\beta_i$ represent the followings: $a_i$ and $b_i$ denote the degradation rates of the $i$-th mRNA and protein, respectively; $c_i$ and $\beta_i$ denote the translation and transcription rates, respectively. The nonlinear function $f_i(\cdot) : \mathbb{R}_+ \to \mathbb{R}_+$ stands for the effect of either activation or repression of the transcription, and it is defined by

$$f_i(p_{i-1}) := \begin{cases} \dfrac{1}{1+p_{i-1}^\nu} (=: F_R(p_{i-1})) & \text{(repression)} \\ \dfrac{p_{i-1}^\nu}{1+p_{i-1}^\nu} (=: F_A(p_{i-1})) & \text{(activation)}, \end{cases} \qquad (2)$$

with a Hill coefficient $\nu$.

We see from (1) that the transfer function from $u_i := f_i(p_{i-1})$ to $p_i$, which describes the dynamics of protein production in each gene, becomes a second order system with time delay. Defining these second order systems as $h_i(s)$ ($i = 1, 2, \cdots, N$), the overall system (1) can be described by Fig. 2 (Left), where a transfer matrix $H(s)$ and a static vector

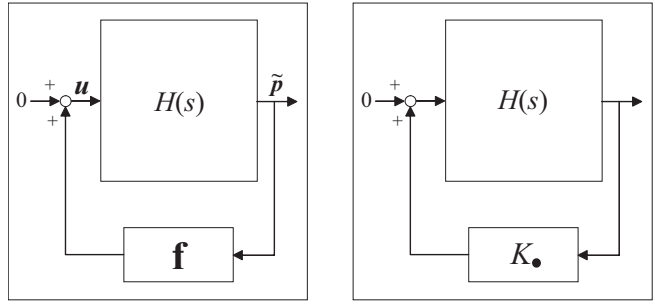

Figure 2: (Left) The block diagram of the gene regulatory networks with negative cyclic feedback, where $\tilde{\boldsymbol{p}} := [p_1(t-\tau_{p_1}), p_2(t-\tau_{p_2}), \cdots, p_N(t-\tau_{p_N})]^T$. (Right) linear system $\mathcal{H}_\bullet(s)$ in (16). The static nonlinearity $\boldsymbol{f}$ is replaced with the corresponding describing function.

nonlinearity function $\mathbf{f}$ are defined as

$$H(s) := \mathrm{diag}(h_1(s), h_2(s), \cdots, h_N(s)), \tag{3}$$

$$\boldsymbol{f} := [R_1^2 f_1(\cdot), R_2^2 f_2(\cdot), \cdots, R_N^2 f_N(\cdot)]^T. \tag{4}$$

The transfer function $h_i(s)$ and the constants $R_i$ are

$$h_i(s) := \frac{e^{-s(\tau_{r_i}+\tau_{p_i})}}{(T_{a_i}s+1)(T_{b_i}s+1)}, \ T_{a_i} := \frac{1}{a_i}, \ T_{b_i} := \frac{1}{b_i}, \tag{5}$$

$$R_i := \frac{\sqrt{c_i \beta_i}}{\sqrt{a_i b_i}}. \ (i=1,2,\cdots,N). \tag{6}$$

It should be emphasized that $H(s)$ describes the dynamics of each gene's protein production, and $\boldsymbol{f}$ describes the interaction between genes.

The dynamical behavior of the system (1) can be qualitatively classified by

$$\delta := \prod_{i=1}^N \delta_i, \ \delta_i := \begin{cases} +1 & (f_i(\cdot) = F_A(\cdot)) \\ -1 & (f_i(\cdot) = F_R(\cdot)) \end{cases}. \tag{7}$$

Specifically, the protein concentrations asymptotically converge to one of equilibria when $\delta > 0$, while they exhibit oscillatory solutions as well as convergence when $\delta < 0$ (Enciso, 2007; Smith, 1987). Therefore, the gene regulatory networks with $\delta < 0$ are of interest in this paper.

**Assumption 1.** *For given $f_i(\cdot)$ ($i=1,2,\cdots,N$), $\delta < 0$.*

This assumption implies that a given gene regulatory network has an odd number of repressive interactions ($df_i/dp < 0$) between genes. Thus, the loop gain of the overall system is negative.

In order to capture the essential dynamical properties in an analytic way, we hereafter consider a simplified model of (1). It is assumed that the kinetic parameters $a_i, b_i, c_i$ and $\beta_i$ in (1) are homogeneous between genes throughout this paper.

**Assumption 2.** $a_1 = a_2 = \cdots = a_N(=:a), b_1 = b_2 = \cdots = b_N(=:b), c_1 = c_2 = \cdots = c_N(=:c)$ and $\beta_1 = \beta_2 = \cdots = \beta_N(:=\beta)$ in (1).

**Remark 1.** In Appendix A, we also discuss the case where the parameters are heterogeneous. In particular, we show that the analysis based on Assumption 2 provides important insights for the case of heterogeneous parameters as well, though the primary aim of Assumption 2 is to simplify the model and explore the essence of the dynamics in an analytic way.

Table 1: Physical meanings of the constants

| | |
|---|---|
| $N$ | The number of genes in gene regulatory network |
| $Q$ | Discrepancy of mRNA and protein degradation time |
| $R$ | Ratio of degradation and production rates, which accounts for equilibrium concentrations |
| $\tilde{\tau}$ | Mean time delay in transcription and translation normalized by the mean degradation rate |
| $\nu$ | Hill coefficient, which quantifies the degree of cooperative binding |

## 2.2 Existence of oscillations

In Takada et al. (2010), the authors presented existence conditions of oscillations for the gene regulatory system (1). The analysis then revealed that five dimensionless quantities essentially determine the existence of oscillations, namely $(N, Q, R, \tilde{\tau}, \nu)$, where

$$Q := \frac{T_G}{T_A}, \ R := \frac{c\beta}{ab}, \ \tilde{\tau} := \frac{\tau}{T_A}, \ T_A := \frac{T_a + T_b}{2}, \ T_G := \sqrt{T_a T_b} \tag{8}$$

with

$$T_a := \frac{1}{a}, \ T_b := \frac{1}{b}, \ \tau := \frac{\sum_{i=1}^{N}(\tau_{r_i} + \tau_{p_i})}{N}. \tag{9}$$

Physical meanings of these constants are summarized in Table 1.

**Remark 2.** Since $Q$ is the ratio of geometric and arithmetic means, $0 < Q \leq 1$ holds. In particular, $Q \to 1$ implies that the degradation time constants of mRNA and proteins, $T_a$ and $T_b$, tend to be a same value. The parameter $\tilde{\tau}$ describes the ratio of mean transcription and translation delay to the mean degradation rate.

In what follows, we analyze the frequency, phase and amplitude of oscillatory protein concentrations, and reveal how these parameters in Table 1 affect the oscillation profiles. Throughout this paper, it is assumed that the system (1) satisfies the existence conditions of oscillations in Takada et al. (2010), and exhibits oscillatory protein concentrations.

# 3 Multivariable Harmonic Balance Analysis for Biochemical Oscillators

In this section, we provide a theoretic framework of oscillation profile analysis for the gene regulatory network (1). Using the harmonic balance analysis (Khalil, 2001; Iwasaki, 2008), we first derive a quasi-linear system associated with (1). Then, it is shown that the frequency, phase and amplitude profiles are obtained from the equations that the closed loop quasi-linear system should satisfy.

## 3.1 Multivariable harmonic balance analysis

The oscillatory waveform of $p_i(t)$ $(i = 1, 2, \cdots, N)$ can be written with the infinite sum of sinusoidal waves as $p_i(t) = \sum_{k=0}^{\infty} \alpha_{ik} \sin(k\varpi t + \varphi_{ik})$ with some constants $\varpi, \varphi_{ik}$ and $\alpha_{ik}$ $(k = 0, 1, \cdots)$. Here, we can expect that $p_i(t)$ is approximately written as

$$p_i(t) \simeq x_i + y_i \sin(\varpi t + \varphi_i) \ (i = 1, 2, \cdots, N), \tag{10}$$

since the higher order harmonic components are attenuated by the second order low-pass filters $h_i(s)$ in the cyclic network. It should be noted that $x_i$ and $y_i$ denote the bias and the amplitude of the first order harmonic components of the $i$-th protein $p_i(t)$, respectively. We assume $x_i \geq 0$ and $y_i \geq 0$ without loss of generality.

Let the describing functions of $R^2 f_i(\cdot)$ be defined by

$$\eta_i(x_{i-1}, y_{i-1}) := \frac{R_i^2}{2\pi x_{i-1}} \int_{-\pi}^{\pi} f_i\left(x_{i-1} + y_{i-1} \sin(t)\right) dt, \tag{11}$$

$$\xi_i(x_{i-1}, y_{i-1}) := \frac{R_i^2}{\pi y_{i-1}} \int_{-\pi}^{\pi} f_i\left(x_{i-1} + y_{i-1} \sin(t)\right) \sin(t) dt. \tag{12}$$

The describing functions $\eta_i(x_{i-1}, y_{i-1})$ and $\xi_i(x_{i-1}, y_{i-1})$ represent the bias and harmonic gains of $R_i^2 f_i(\cdot)$ for the sinusoidal input $x_{i-1} + y_{i-1} \sin(\varpi t)$, respectively (Khalil, 2001).

Approximating the nonlinearity $f(\cdot)$ with the describing functions, we see that the input to $h_i(s)$ is $u_i = \eta_i(x_{i-1}, y_{i-1})$ and $u_i = \xi_i(x_{i-1}, y_{i-1})$ instead of $u_i = f_i(\cdot)$. Thus, the system depicted in Fig. 2 (Left) can be redrawn as shown in Fig. 2 (Right), where

$$\begin{aligned} \mathcal{K}_0(\boldsymbol{x}, |\boldsymbol{y}|) &:= \operatorname{cyc}(\eta_1, \eta_2, \cdots, \eta_N), \\ \mathcal{K}_1(\boldsymbol{x}, |\boldsymbol{y}|) &:= \operatorname{cyc}(\xi_1, \xi_2, \cdots, \xi_N), \end{aligned} \tag{13}$$

and

$$\operatorname{cyc}(z_1, z_2, z_3, \cdots, z_N) := \begin{bmatrix} 0 & 0 & 0 & \cdots & z_1 \\ z_2 & 0 & 0 & \ddots & 0 \\ 0 & z_3 & 0 & \ddots & \vdots \\ \vdots & \ddots & \ddots & \ddots & \vdots \\ 0 & \cdots & 0 & z_N & 0 \end{bmatrix}.$$

The symbols $\boldsymbol{x}$ and $\boldsymbol{y}$ are defined as $\boldsymbol{x} := [x_1, x_2, \cdots, x_N]^T \in \mathbb{R}_+^N$ and $\boldsymbol{y} := [y_1 e^{j\tilde{\varphi}_1}, y_2 e^{j\tilde{\varphi}_2}, \cdots, y_N e^{j\tilde{\varphi}_N}]^T \in \mathbb{C}^N$ with

$$\tilde{\varphi}_i := \varphi_i - \varpi \tau_{p_i}, \tag{14}$$

and $|\boldsymbol{y}|$ stands for elementwise absolute values, i.e., $|\boldsymbol{y}| = [y_1, y_2, \cdots, y_N]^T \in \mathbb{R}_+^N$.

We can see that the system in Fig. 2 (Right) would satisfy the following closed-loop equations, if the waveform of $p_i(t)$ was strictly the sinusoidal wave $x_i + y_i \sin(\varpi t + \varphi_i)$.

$$(I - H(0)\mathcal{K}_0(\boldsymbol{x}, |\boldsymbol{y}|))\boldsymbol{x} = 0 \tag{15a}$$
$$(I - H(j\varpi)\mathcal{K}_1(\boldsymbol{x}, |\boldsymbol{y}|))\boldsymbol{y} = 0. \tag{15b}$$

Thus, the solution $(\varpi, \boldsymbol{x}, \boldsymbol{y})$ of the equations (15) is expected to capture an approximate profile of the oscillations when $p_i(t)$ is sufficiently close to the sinusoidal of the form (10). Consequently, the oscillation profile analysis reduces to the problem of finding $3N$ variables $(\varpi, x_1, x_2, \cdots, x_N, y_1, y_2, \cdots, y_N, \varphi_2, \varphi_3, \cdots, \varphi_N)$ satisfying (15). Note that $\varphi_1$ is defined as $\varphi_1 := 0$ without loss of generality.

Let $\boldsymbol{x}^* = [x_1^*, x_2^*, \cdots, x_N^*]^T \in \mathbb{R}^N$ and $\boldsymbol{y}^* = [y_1^* e^{j\tilde{\varphi}_1}, y_2 e^{j\tilde{\varphi}_2^*}, \cdots, y_N^* e^{j\tilde{\varphi}_N^*}]^T \in \mathbb{C}^N$ denote the constant vectors that satisfy both of (15). We define the linear time-invariant systems $\mathcal{H}_0(s)$ and $\mathcal{H}_1(s)$ as

$$\mathcal{H}_\bullet(s) := (I - H(s)K_\bullet)^{-1} \quad (\bullet = 0, 1), \tag{16}$$

where $K_\bullet$ is the constant matrices defined by $K_\bullet := \mathcal{K}_\bullet(\boldsymbol{x}^*, |\boldsymbol{y}^*|)$ ($\bullet = 0, 1$). Although there may exist multiple solutions for the equations (15), orbitally unstable solutions can be empirically ruled out by checking the stability of $\mathcal{H}_\bullet(s)$ (see Khalil (2001); Iwasaki (2008) and references therein). Specifically, a pair of poles of $\mathcal{H}_\bullet(s)$ is expected to lie on the imaginary axis, and the rest in the open left half complex plane, if a solution ($\varpi$, $\boldsymbol{x}^*$, $\boldsymbol{y}^*$) is orbitally stable. Thus, the profiles of stable oscillations can be specified from the solutions of (15) and the marginal stability condition.

### 3.2 Oscillation profile analysis for biochemical oscillators

Using the harmonic balance approach, the oscillation profile analysis of biochemical oscillators can be summarized as follows.

> **Oscillation profile analysis:** Consider the gene regulatory network modeled by (1). Suppose $p_i(t)$ is approximately written as $p_i(t) \simeq x_i + y_i \sin(\varpi t + \varphi_i)$ ($i = 1, 2, \cdots, N$). Then, we find ($\varpi, \varphi_i, x_i, y_i$) such that the following two conditions hold.
>
> **(C1)** The equations (15a) and (15b) hold.
>
> **(C2)** The system $\mathcal{H}_0(s)(\mathcal{H}_1(s))$ has a pole at $s = 0 (s = \pm j\varpi)$, and the rest in the open left half complex plane.

In what follows, we explore the solutions of (15) in an analytic way. We show that the equations (15) can be rewritten in the form of eigen-equations, and the profiles of oscillations can be determined from the eigenvalues/eigenvectors of matrices with a certain structure.

Let a $N \times N$ diagonal transfer matrix $U(s)$ and a scalar transfer function $h(s)$ be defined by

$$U(s) := \mathrm{diag}(e^{s(\tau - \tau_1)}, e^{s(\tau - \tau_2)}, \cdots, e^{s(\tau - \tau_N)}), \tag{17}$$

$$h(s) := \frac{1}{(T_a s + 1)(T_b s + 1)}, \tag{18}$$

where

$$\tau_i := \tau_{r_i} + \tau_{p_i} \ (i = 1, 2, \cdots, N). \tag{19}$$

It follows that $H(s) = h(s)e^{-s\tau}U(s)$ with $\tau$ defined in (9). Note that $H(s)$ is decomposed into the common dynamics $h(s)e^{-s\tau}$ and the deviation $U(s)$. Dividing (15) by $h(0)$ and $h(j\varpi)e^{-j\varpi}$ and substituting the solution ($\varpi, \boldsymbol{x}^*, \boldsymbol{y}^*$), we have

$$(\phi(0)I - K_0)\boldsymbol{x}^* = 0, \tag{20a}$$

$$(\phi(j\varpi)e^{j\varpi\tau}I - UK_1)\boldsymbol{y}^* = 0. \tag{20b}$$

where

$$\phi(s) := 1/h(s) \text{ and } U := U(j\varpi). \tag{21}$$

We see that $\boldsymbol{x}^*$ and $\boldsymbol{y}^*$ can be seen as the eigenvectors of $K_0$ and $UK_1$ associated with the eigenvalues $\phi(0)$ and $\phi(j\varpi)e^{j\varpi\tau}$, respectively. However, the solution is not obtained by simply analyzing the eigenvalue and the associated eigenvector. This is because the matrices $K_0, K_1$ and $U$ depend on $\boldsymbol{x}, |\boldsymbol{y}|$ and $\varpi$ as shown in (13) and (21), thus we need to determine both the matrices $\mathcal{K}_\bullet(\boldsymbol{x}, |\boldsymbol{y}|)$ and the corresponding eigenvalues/eigenvectors simultaneously.

Nevertheless, this observation provides us with important intuitions for our subsequent analysis. In the following sections, we show that (20) can be analytically solved by using the structure of the matrices $\mathcal{K}_\bullet(\boldsymbol{x}, |\boldsymbol{y}|)$.

# 4 Profiles of Oscillations

In this section, we analytically derive the profiles of the oscillatory protein concentrations in terms of the biological parameters in Table 1.

## 4.1 The frequency of oscillations

We first obtain the frequency $\varpi$ in terms of the parameters shown in Table 1. The following lemma describes the eigenvalue distribution of $U(j\varpi)\mathcal{K}_1(\boldsymbol{x}, |\boldsymbol{y}|)$.

**Lemma 1.** *For any given $(\varpi, \boldsymbol{x}, \boldsymbol{y})$, the eigenvalues $\lambda_i$ ($i = 1, 2, \cdots, N$) of the matrix $U(j\varpi)\mathcal{K}_1(\boldsymbol{x}, |\boldsymbol{y}|)$ are given by*

$$\lambda_i := \left| \prod_{k=1}^{N} \xi_k(x_{k-1}, y_{k-1}) \right|^{\frac{1}{N}} e^{j\frac{2i-1}{N}\pi}. \tag{22}$$

**Proof.** For any given $(\varpi, \boldsymbol{x}, \boldsymbol{y})$, the characteristic equation of $U(j\varpi)\mathcal{K}_1(\boldsymbol{x}, |\boldsymbol{y}|)$ is written as

$$s^N - \prod_{i=1}^{N} \xi(x_{i-1}, y_{i-1}) = 0. \tag{23}$$

Note that the time delays are cancelled out. It follows that $\prod_{i=1}^{N} \xi(x_{i-1}, y_{i-1}) < 0$, because $\xi_i(x_{i-1}, y_{i-1})$ is negative/positive when $\delta_i$ is negative/positive, and Assumption 1 holds. Thus, (22) is derived by solving (23). □

We see that the eigenvalues are equiangularly located on a circle with center at the origin. Moreover, $\varpi$ does not affect the eigenvalue, and $\boldsymbol{x}$ and $\boldsymbol{y}$ affect only the radial position, but not the the angular position of the eigenvalues. Since the oscillation frequency $\varpi$ should satisfy (20), possible candidates of $\varpi$ can be characterized as

$$\varpi \in \left\{ \omega \;\middle|\; \arg(\phi(j\omega)e^{j\omega\tau}) = \left\{\frac{2i-1}{N}\right\}_{i=1}^{N} \right\}, \tag{24}$$

where $\arg(\cdot)$ stands for the argument of a complex number. Note that (24) does not depend on $\boldsymbol{x}$ and $\boldsymbol{y}$. The possible solutions of $\varpi$ are illustrated in Fig. 3. We see from Fig. 3 that there are countably infinite solutions.

Despite the infinite candidates of $\varpi$, we can show that an orbitally stable solution is unique by considering the condition (C2). The following lemma gives a necessary and sufficient condition for marginal stability of $\mathcal{H}_1(s)$.

**Lemma 2.** *The system $\mathcal{H}_1(s)$ defined by (16) has a pair of poles at $s = \pm j\omega_0$ and the rest in the open left half complex plane, if and only if a pair of the eigenvalues of $UK_1$, which we denote by $\lambda_\ell$ and $\lambda'_\ell (= \bar{\lambda}_\ell)$, satisfies $\phi(j\omega_0) = \lambda_\ell$ and $\phi(-j\omega_0) = \lambda'_\ell$, and the others lie in the domain $\Omega^c_+ := \{\gamma \in \mathbb{C} \mid \phi(s)e^{s\tau} \neq \gamma \text{ for } \forall s \in \mathbb{C}_+\}$.*

The proof of Lemma 2 is given in Appendix B. The domain $\Omega^c_+$ is the hatched region partitioned by $\phi(j\omega)e^{j\omega\tau}$ in Fig. 3. Lemma 2 implies that the marginal stability of $\mathcal{H}_1(s)$ is examined from the domain $\Omega^c_+$ and the eigenvalue distribution of $UK_1$. In particular, the gain and phase monotonicity of $\phi(s)e^{s\tau}$ allows us to show the uniqueness of the frequency $\varpi$ that satisfies both (C1) and (C2).

**Proposition 1.** *Suppose there exist $(\varpi, \boldsymbol{x}, \boldsymbol{y})$ satisfying (C1) and (C2). Then, the frequency $\varpi$ is uniquely given by the minimum positive solution of $\arg(\phi(j\varpi)e^{j\varpi\tau}) = \pi/N$.*

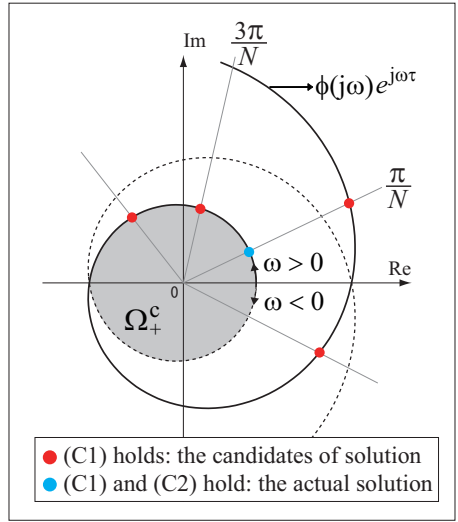

Figure 3: Graphical interpretation of the harmonic balance equation. The red points satisfy the bias and the harmonic balance equations, but does not satisfy the marginal stability condition.

**Proof.** The distance of the boundary of $\Omega_+^c$ from the origin, *i.e.*,

$$|\phi(j\omega)e^{j\omega\tau}| = \sqrt{(1+T_a\omega^2)(1+T_b\omega^2)}$$

monotonically increases for $\omega \geq 0$. On the other hand, the eigenvalues of $UK_1$ are equally distant from the origin as seen in Lemma 1. Thus, the marginal stability condition in Lemma 2 implies that $\lambda_1$ in (22) must lie on the boundary of $\Omega_+^c$, when $\mathcal{H}_1(s)$ is marginally stable (see Fig. 3). Consequently, $\varpi$ satisfies $\phi(j\varpi)e^{j\varpi\tau} = \lambda_1$. □

From this proposition, we can analytically derive the frequency profile in terms of the biological parameters.

**Theorem 1.** *Consider the gene regulatory networks modeled by (1). The frequency $\varpi$ of the oscillatory protein concentrations $p_i(t)(i = 1, 2, \cdots, N)$ is expected to be the minimum positive solution of*

$$\varpi = \frac{1}{Q^2}\left(\sqrt{\cot^2\left(\frac{\pi}{N} - \varpi\tau\right) + Q^2} - \cot\left(\frac{\pi}{N} - \varpi\tau\right)\right)\frac{1}{T_A}. \qquad (25)$$

**Proof.** The real and imaginary part of $\phi(j\omega)$ can be written as

$$\text{Re}[\phi(j\omega)] = 1 - T_aT_b\omega^2, \qquad (26)$$
$$\text{Im}[\phi(j\omega)] = (T_a + T_b)\omega. \qquad (27)$$

Note that $\phi(j\omega)$ ($\omega \in \mathbb{R}$) draws a parabolic curve on the complex plane. Proposition 1 implies that $\varpi$ is the minimum positive solution of $\arg(\phi(j\varpi)) = \pi/N - \varpi\tau$. Thus, $\text{Re}[\phi(j\varpi)] = A\cos(\pi/N)$ and $\text{Im}[\phi(j\varpi)] = A\sin(\pi/N)$ for some $A > 0$. Substituting these into (26) and (27), and eliminating $A$, we have the following equation.

$$T_aT_b\varpi^2 + (T_a + T_b)\cot\left(\frac{\pi}{N} - \varpi\tau\right)\varpi - 1 = 0, \qquad (28)$$

from which (25) is immediately obtained. □

The oscillation frequency is analytically predicted in this theorem. Thus, it is possible to interpret the relation between the parameters and the frequency. Note that (25) is written only in terms of the essential parameters shown in Table 1. We will demonstrate biological insights obtained from Theorem 1 in Section 6.

**Remark 3.** The minimum positive solution of (25) can be efficiently obtained by bisection search for $\varpi \in [0, \pi/N\tau]$, because the right-hand side of (25) monotonically decreases for $\varpi \in [0, \pi/N\tau]$, and the minimum positive solution of (25) exists in this region. It should be noted that $\varpi$ is obtained without the numerical computation as

$$\varpi = \frac{1}{Q^2}\left(\sqrt{\cot^2\left(\frac{\pi}{N}\right)+Q^2}-\cot\left(\frac{\pi}{N}\right)\right)\frac{1}{T_A}. \tag{29}$$

when the time delays are not considered, i.e., $\tau_{r_i} = 0$ and $\tau_{p_i} = 0$ ($i = 1, 2, \cdots, N$).

## 4.2 The phase of oscillations

The phase profile is determined from the eigenvector $\boldsymbol{y}^*$ of $UK_1$ in (20b). In what follows, the phase profile is explored in an analytic approach.

We first show that $UK_1$ can always be transformed into a circulant matrix (Davis, 1979) by similarity transformation.

**Lemma 3.** Let $D := \mathrm{diag}(d_1, d_2, \cdots, d_N) \in \mathbb{C}^{N\times N}$ be defined by

$$d_i := \begin{cases} (-1)^{i-1}\dfrac{\prod_{k=1}^{i}\xi_k^* e^{j\varpi(\tau-\tau_k)}}{\left|\prod_{k=1}^{N}\xi_k^*\right|^{\frac{i-1}{N}}} & (\text{if } N \text{ is odd}) \\[2ex] \dfrac{\prod_{k=1}^{i}\xi_k^* e^{j\varpi(\tau-\tau_k)}}{\left|\prod_{k=1}^{N}\xi_k^*\right|^{\frac{i-1}{N}}}e^{-j\frac{i}{N}\pi} & (\text{if } N \text{ is even}) \end{cases} \tag{30}$$

with $\xi_i^* := \xi_i(x_{i-1}^*, y_{i-1}^*)$. Then, $D^{-1}(UK)D = V$, where $V$ is the circulant matrix of the form

$$V := \begin{cases} |\prod_{k=1}^{N}\xi_k^*|^{\frac{1}{N}}\mathrm{cyc}(-1, -1, \cdots, -1) & (\text{if } N \text{ is odd}) \\ |\prod_{k=1}^{N}\xi_k^*|^{\frac{1}{N}}\mathrm{cyc}(e^{\frac{j\pi}{N}}, e^{\frac{j\pi}{N}}, \cdots, e^{\frac{j\pi}{N}}) & (\text{if } N \text{ is even}). \end{cases}$$

Since circulant matrices are known to be diagonalized by the discrete Fourier transform matrix (Davis, 1979), the eigenvector $\boldsymbol{q} := [q_1, q_2, \cdots, q_N]^T$ of $V$ associated with the eigenvalue $\phi(j\varpi)e^{j\varpi\tau}$ is easily obtained as

$$q_i := \begin{cases} (-1)^i e^{-j\frac{i-1}{N}\pi} & (\text{if } N \text{ is odd}) \\ 1 & (\text{if } N \text{ is even}). \end{cases} \tag{31}$$

Therefore, the phasor $\boldsymbol{y} = [y_1 e^{j\tilde{\varphi}_1}, y_2 e^{j\tilde{\varphi}_2}, \cdots, y_N e^{j\tilde{\varphi}_N}]^T$ is calculated from $\boldsymbol{y} = D\boldsymbol{q}$. Finally, computing $\varphi_i = \tilde{\varphi}_i + \varpi\tau_{p_i}$ ($i = 1, 2, \cdots, N$) yields the following analytic estimate of the phase.

**Theorem 2.** Consider the gene regulatory networks modeled by (1). The phase shift ($\varphi_{i+1} - \varphi_i$) between the $(i+1)$-th and the $i$-th protein is expected as

$$\varphi_{i+1} - \varphi_i = \left(Z_i - \frac{1}{N}\right)\pi - \varpi\Delta\tau_i. \tag{32}$$

for $i = 1, 2, \cdots, N$, where

$$\Delta \tau_i := (\tau_{r_{i+1}} + \tau_{p_i}) - \tau, \tag{33}$$

$$Z_i := \begin{cases} 1 & \text{if } \delta_{i+1} = -1 \\ 0 & \text{if } \delta_{i+1} = +1 \end{cases}, \tag{34}$$

and $\varpi$ is given by Theorem 1.

This theorem analytically predicts the phase difference between protein species. The constants $\Delta \tau_i$ $(i = 1, 2, \cdots, N)$ are the discrepancy of the time delay of the $i$-th gene from the average delay $\tau$. The interpretation of Theorem 2 is given in Section 6.

**Remark 4.** When the delays are homogeneous between genes, i.e., $\tau_{r_1} = \tau_{r_2} = \cdots = \tau_{r_N}$ and $\tau_{p_1} = \tau_{p_2} = \cdots = \tau_{p_N}$, (32) is independent of $\varpi$, and it depends only on the activation-repression pattern of the regulatory network.

### 4.3 The bias and amplitude of oscillations

The bias and amplitude profiles can be predicted from $\boldsymbol{x}$ and $|\boldsymbol{y}|$ in (15).

We first show that the eigenvalues and eigenvectors of $\mathcal{K}_0(\boldsymbol{x}, |\boldsymbol{y}|)$ are obtained in a similar fashion to Lemma 1.

**Lemma 4.** *For any given $(\boldsymbol{x}, \boldsymbol{y})$, the eigenvalues $\mu_i$ $(i = 1, 2, \cdots, N)$ of the matrix $\mathcal{K}_0(\boldsymbol{x}, |\boldsymbol{y}|)$ are given by*

$$\mu_i := \left| \prod_{k=1}^{N} \eta_k(x_{k-1}, y_{k-1}) \right|^{\frac{1}{N}} e^{j \frac{2(i-1)}{N} \pi}, \tag{35}$$

*and the eigenvector associated with $\mu_1$ is $[\eta_1, \eta_1\eta_2, \cdots, \prod_{i=1}^{N} \eta_i]^T$.*

This lemma shows that the eigenvalues of $K_0$ are located on a circle with center at the origin, but the angular position is different from that of $UK_1$ (see Lemma 1). Following the similar argument as Lemma 1 and Proposition 1, we can show that $\mu_1$ should become the eigenvalue that corresponds to $\phi(0)(= 1)$ in (20a). Therefore, the bias $\boldsymbol{x}^*$ is obtained from the eigenvector of $K_0$ associated with $\mu_1$, and we have the following theorem.

**Theorem 3.** *Consider the gene regulatory networks modeled by (1). The bias $\boldsymbol{x}$ and amplitude $|\boldsymbol{y}|$ of oscillatory protein concentrations $p_i(t)$ $(i = 1, 2, \cdots, N)$ are expected to satisfy both (36a) and (36b) simultaneously.*

$$\frac{x_{i+1}}{x_i} = \eta_{i+1}(x_i, y_i) \tag{36a}$$

$$\frac{y_{i+1}}{y_i} = \frac{\xi_{i+1}(x_i, y_i)}{|\xi_k(x_{k-1}, y_{k-1})|^{\frac{1}{N}}} = \frac{\xi_{i+1}(x_i, y_i)}{|\phi(j\varpi)|}, \tag{36b}$$

*where $\varpi$ is given by Theorem 1.*

This theorem provides the equations that the bias and amplitude of oscillations should satisfy. Note that $x_i$ and $y_i$ $(i = 1, 2, \cdots, N)$ are determined from the recursive equations (36) once $(x_1, y_1)$ is determined. It also follows that

$$\prod_{i=1}^{N} \eta_i(x_{i-1}, y_{i-1}) = 1 \tag{37a}$$

$$\left| \prod_{i=1}^{N} \xi_i(x_{i-1}, y_{i-1}) \right|^{\frac{1}{N}} = |\phi(j\varpi)|, \tag{37b}$$

from (20). Thus, $x_i$ and $y_i$ ($i = 1, 2, \cdots, N$) could be obtained by numerically searching $(x_1, y_1)$ so that (37) is satisfied under the constraint (36). However, the construction of a fast and reliable algorithm to compute $\boldsymbol{x}$ and $|\boldsymbol{y}|$ remains a future challenge.

## 5 Applications to Biochemical Oscillators

In this section, we demonstrate the proposed analysis method with two biochemical networks. We first consider the Pentilator (Tsai et al., 2008), a conceptual biochemical network consisting of $N = 5$ genes, to illustrate that the developed result is useful even for the large network. Then, we investigate the oscillations of an existing oscillatory reactions, the somitogenesis clock (Hirata et al., 2004).

### 5.1 Pentilator with time delay

Pentilator (Tsai et al., 2008) is a model of the gene regulatory network that is composed of $N = 5$ genes interacting in a cyclic way. Although the original model assumed that all the interactions are repressive, we here replace two of the repressors with activators in order to describe the effect of activators on the oscillation profiles.

We consider the gene regulatory network depicted in Fig. 4 (Top). The dynamical model of the Pentilator in Fig. 4 (Top) is then written for $i = 1, 2, \cdots, 5$ as

$$\begin{aligned} \dot{r}_i(t) &= -ar_i(t) + \beta f_i(p_{i-1}(t - \tau_{p_{i-1}})), \\ \dot{p}_i(t) &= cr_i(t - \tau_{r_i}) - bp_i(t), \end{aligned} \tag{38}$$

where $f_1(\cdot) = f_2(\cdot) = f_4(\cdot) = F_R(\cdot)$ and $f_3(\cdot) = f_5(\cdot) = F_A(\cdot)$. Note that time delays of transcription and translation are explicitly modeled in (38), though they were not introduced in Tsai et al. (2008).

The parameters are set as follows: the degradation rates are $a = 2.0[\text{min}^{-1}]$ and $b = 0.2[\text{min}^{-1}]$, and the synthesis rates are $c = 0.3[\text{min}^{-1}]$ and $\beta = 10[\text{min}^{-1}]$. The Hill coefficient is $\nu = 2.0$, and the time delays are $\boldsymbol{\tau}_r := [1.8, 1.4, 1.1, 0.7, 1.0]^T[\text{min}]$ and $\boldsymbol{\tau}_p := [1.0, 0.8, 0.4, 0.4, 0.4]^T[\text{min}]$, where the $i$-th entry of $\boldsymbol{\tau}_r$ and $\boldsymbol{\tau}_p$ is defined as $\tau_{r_i}$ and $\tau_{p_i}$, respectively.

From the above parameters, we have

$$\begin{aligned} Q &= 0.575, \ \tau = 1.8, \ T_A = 1.1, \\ \Delta\boldsymbol{\tau} &= [1.0, 0.4, -0.3, -0.7, -0.4]^T, \end{aligned} \tag{39}$$

where $\Delta\boldsymbol{\tau}$ is the vector whose $i$-th entry is $\Delta\tau_i$ defined in (33). Then, $\phi(s)e^{s\tau}$ in (20) becomes $\phi(s)e^{s\tau} = (5s+1)(0.5s+1)e^{1.8s}$, and the problem reduces to finding $(\varpi, \boldsymbol{x}, \boldsymbol{y})$ satisfying the conditions (C1) and (C2).

The existence of oscillations for this system can be confirmed from the conditions presented in Takada et al. (2010), thus we hereafter analyze the waveform of the oscillations. Using Theorem 1, we first obtain the oscillation frequency as $\varpi = 8.98 \times 10^{-2}[\text{rad/min}]$. The phase profiles can be computed from Theorem 2, where $Z_1 = -1, Z_2 = -1, Z_3 = +1, Z_4 = -1, Z_5 = +1$. The results are summarized in Table 2. Note that we define $\varphi_1 = 0$ without loss of generality.

To confirm these results, we conducted a numerical simulation of (38). The time course of protein concentrations is presented in Fig. 4 (Bottom). The frequency and phase of the simulated oscillations are shown in Table 2. We see that the proposed analytic results provide approximate oscillation profiles with high accuracy.

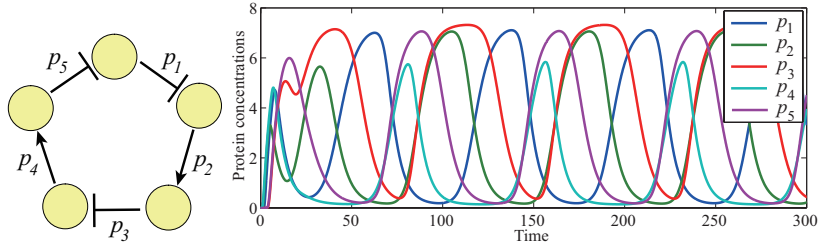

Figure 4: (Top) Schematic diagram of the rearranged Pentilator considered in this example. (Bottom) Time course of the oscillatory protein concentrations in the rearranged Pentilator.

Table 2: The estimated/actual frequency and phase of the oscillations

| Theorem 1 | $8.98 \times 10^{-2}$ [rad/min] |
|---|---|
| Simulation | $8.61 \times 10^{-2}$ [rad/min] |
| Relative error | 4.30 % |

|  | $p_2$ | $p_3$ | $p_4$ | $p_5$ |
|---|---|---|---|---|
| Theorem 2 [deg] | 146.1 | 108.5 | 255.6 | 218.1 |
| Simulation [deg] | 141.1 | 110.1 | 251.7 | 219.8 |

## 5.2 Somitogenesis clock

Somitogenesis is the process by which the somites of living organisms are produced. Recent biological studies revealed that somite segmentation of mouse embryos occurs every two hours, and the segmentation clock is primarily controlled by oscillatory gene expression of *Hes7* gene (see Hirata et al. (2004) and references therein).

The *Hes7* transcription is controlled by the self-negative feedback of Hes7 protein. Thus, the dynamics of Hes7 protein concentration can be described as

$$\begin{cases} \dot{r}_1(t) = -ar_1(t) + \dfrac{\beta}{1 + p_1^2(t - \tau_{p_1})}, \\ \dot{p}_1(t) = -bp_1(t) + cr_1(t - \tau_{r_1}). \end{cases} \qquad (40)$$

Note that this model is a special case of (1) with $N = 1$.

Theoretical studies of the self-negative feedback system (40) predicted that short half-life of Hes7 is a key to generate the oscillations of two-hour period (Monk, 2003) . Later, Hirata et al. (2004) conducted an experiment using Hes7 mutants with long half-lives and normal repressor activity, and confirmed that the mutants with long half-lives do not exhibit oscillations. Although the oscillations' existence was focused in many existing studies including the authors' work (Takada et al., 2010), the underlying mechanism that determines the profiles of oscillations are still unclear.

In what follows, we demonstrate that the frequency profile can be analyzed with Theorem 1, then provide insights on the relation between the parameters and the frequency of the oscillations. Following Hirata et al. (2004), we here define the mRNA and protein degradation rates $a$ and $b$ as

$$a = \frac{\ln 2}{t_r}, b = \frac{\ln 2}{t_p}, \qquad (41)$$

where $t_r$ and $t_p$ denote mRNA and protein half-life time. The parameter values of wild-type Hes7 is also taken from Hirata et al. (2004): $t_p = 20$ [min], $t_r = 3$ [min], $a = 0.231$ [min$^{-1}$], $b =$ $0.0347\,[\text{min}^{-1}]$, $c = 4.5\,[\text{min}^{-1}], \beta = 0.825\,[\text{min}^{-1}], \tau_p = 17\,[\text{min}], \tau_r = 20\,[\text{min}], \nu = 2$. Note that mRNA and protein concentrations are normalized by the half maximal effective concentration of the protein, or the constant $p_{\text{crit}}$ in Hirata et al. (2004).

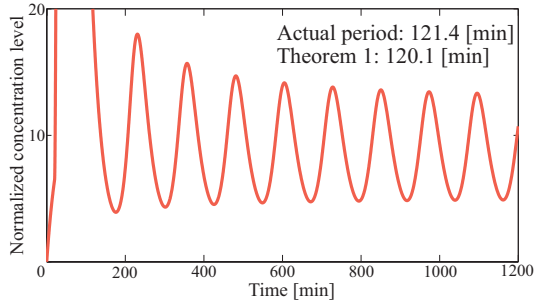

Figure 5: Time trajectory of Hes7 protein concentrations. Hes7 exhibits periodic oscillations with the two-hour period.

Theorem 1 implies that the frequency profile is essentially determined from $Q, \tau$ and $T_A$, whose meanings are listed in Table 1. Thus, these quantities are calculated from the parameters as

$$Q = 0.674, \ \tau = 2.23, T_A = 16.6. \tag{42}$$

We obtain the frequency $\varpi$ from Theorem 1 as $\varpi = 0.523$ [rad/min], from which the period is calculated as

$$\frac{2\pi}{\varpi} = 120.1 \,[\text{min}]. \tag{43}$$

The numerical simulation of Hes7 concentrations indeed exhibits two-hour oscillations as shown in Fig. 5. We see that Theorem 1 successfully predicted the frequency of oscillations without simulating (40).

**Remark 5.** Hirata et al. (2004) showed that the period of oscillations is approximately given by $2(\tau_r + \tau_p + \ln(2)/a + \ln(2)/b)$ with $a$ and $b$ defined by (41). This estimate provides 120.0 [min] in the above example. A more detailed comparison of Theorem 1 and the estimate in Hirata et al. (2004) is shown in Appendix C. It should be emphasized that Theorem 1 is useful for negative cyclic feedback networks consisting of any number of genes, while Hirata's estimate is applicable only for the self-repression case, *i.e.,* $N = 1$.

Theorem 1 also allows us to obtain more general insights on the frequency profile. The equation (25) implies that the parameters $\tau$ and $Q$ are important for characterizing the frequency. Based on Theorem 1, the relation between these parameters and the period $2\pi/\varpi$ of oscillations is illustrated in Fig. 6. Note that $Q$ stands for the discrepancy of mRNA and protein degradation rates, and $Q = 1.0$ implies $a = b = 0.0604[\text{min}^{-1}]$. We see that the frequency is sensitive to the sum of transcription and translation delay $\tau = \tau_r + \tau_p$ rather than $Q$. Thus, the transcription and translation delays play an important role in regulating the two-hour cycle of the segmentation clock.

# 6 Biological Insights

In this section, we provide general biological insight based on Theorem 1 and Theorem 2. Specifically, we illustrate how the profiles of oscillations depend on parameters and structures of the network.

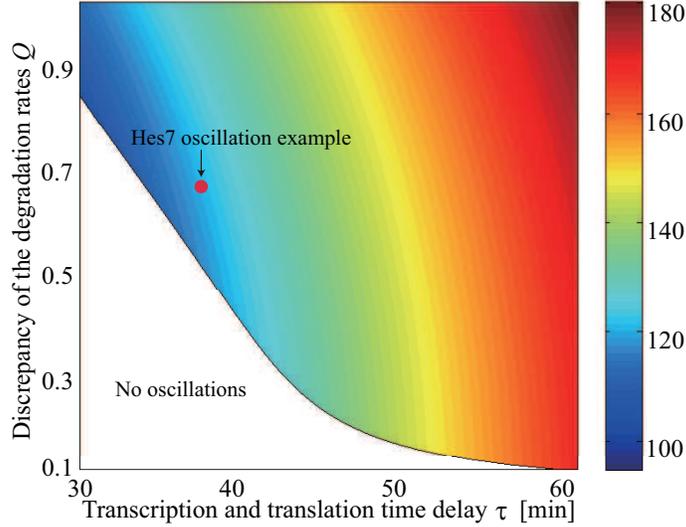

Figure 6: The period of Hes7 oscillations predicted from Theorem 1. The mean degradation rate of mRNA and proteins is set as $T_A = 16.6$, and $R = 21.5$, which is equivalent to the parameters in (39). The oscillations illustrated in Fig. 5 correspond to the red point with the caption "Hes7 oscillation example."

**Frequency:** The frequency profile is given by (25) in Theorem 1. Let $\tilde{\varpi}$ be defined by $\tilde{\varpi} := \varpi T_A$ with $T_A$ in (8). From the definition, $\tilde{\varpi}$ is a frequency normalized by the mean degradation rates of mRNA and proteins. Then, we can rewrite (25) as

$$\tilde{\varpi} = \frac{1}{Q^2}\left(\sqrt{\cot^2\left(\frac{\pi}{N} - \tilde{\varpi}\tilde{\tau}\right) + Q^2} - \cot\left(\frac{\pi}{N} - \tilde{\varpi}\tilde{\tau}\right)\right), \tag{44}$$

where $\tilde{\tau}$ is the time delay normalized by $T_A$ (see (8) for the definition). This implies that time can be normalized by the mean degradation rates $T_A$ without loss of generality. Therefore, the frequency profile can essentially be captured by the dimensionless quantities that appear in (44).

We see that $\tilde{\varpi}$ depends only on the normalized time delay $\tilde{\tau}$, the number of genes $N$ and the discrepancy of the degradation rates $Q$ (see Table 1 for the biological meanings). It is worth noting that as a result of cyclic feedback, the mean time delay over all genes, $\tilde{\tau}$, plays a decisive role rather than the individual delays at each gene. Thus, the frequency can be less sensitive to the variation of the delays in intermediate reactions.

Looking more closely into (44), we obtain more quantitative relations between these parameters and the frequency as follows: the normalized frequency $\tilde{\varpi}$ monotonically increases as

(i) the average time delay of transcription and translation process ($\tilde{\tau}$) decreases

(ii) the number of genes ($N$) decreases

(iii) the mRNA and protein degradation time gets close to each other.

These insights are confirmed by numerical simulations as shown in Fig. 7. Note that the vertical axis is the period of oscillations, where the time is normalized by $T_A$. We see that the solution of (44) successfully predicts the oscillation profiles.

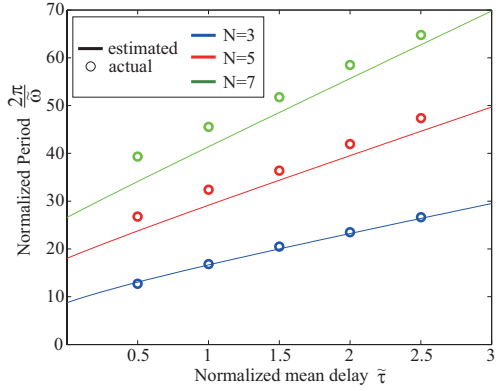

Figure 7: The normalized period $2\pi/\tilde{\varpi}$ of oscillations in terms of $\tilde{\tau}$.

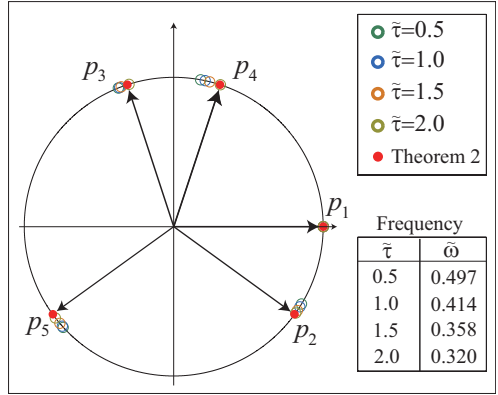

Figure 8: The phase profiles of oscillations for various $\tilde{\tau}$. When $\Delta\boldsymbol{\tau} = \mathbf{0}$, the phase stays intact, while the frequency changes drastically.

In particular, it is observed that the period is almost linearly proportional to the mean delay $\tilde{\tau}$. This observation is essentially consistent with the ones in Jensen et al. (2003); Monk (2003) and Wang et al. (2004), where the period of oscillations was shown to be proportional to the total time delay $\sum_{i=1}^{N}(\tau_{r_i} + \tau_{p_i})$ based on numerical simulations. In fact, we have

$$\frac{2\pi}{\tilde{\varpi}} \simeq 4N + 2\sum_{i=1}^{N}(\tau_{r_i} + \tau_{p_i})\frac{1}{T_A}. \tag{45}$$

by approximating (44) with $\sqrt{1+x^2} \simeq 1+x^2/2$ and $\cot(x) \simeq x^{-1}$, and the normalized period of oscillations $2\pi/\tilde{\varpi}$ is indeed proportional to the sum of delays in Fig. 7 (see Appendix D for the details of the approximation).

**Remark 6.** The frequency profile obtained from (44) becomes less accurate, when the parameters $R$ and $\nu$ are so large that the oscillation waveforms are saturated, because the analysis is based on (10). However, the authors have observed by numerical simulations that the qualitative insights obtained above hold, even when the waveform is much distorted from (10).

**Phase:** The phase of oscillations is given by (32). Since $\varpi\Delta\tau = \tilde{\varpi}\Delta\tilde{\tau}$ with $\Delta\tilde{\tau}_i := \Delta\tau_i/T_A$, (32) can be rewritten with the dimensionless quantities shown in Table 1. We see that the phase is determined from the activation and repression pattern $Z_i$ of the network and $\tilde{\varpi}\Delta\tilde{\tau}_i$.

The constants $\Delta \tilde{\tau}_i$ are the discrepancy of the delay from the mean delay $\tau$. Thus, the delays have no impact on the phase profile if they are homogeneous, while they do have an impact for the frequency profile. An example is illustrated in Fig. 8, where $\Delta \boldsymbol{\tau} = 0$. We see that only the frequency is affected by the difference of the mean delay $\tilde{\tau}$.

The phase profile can be summarized as follows.

(i) The difference of the individual time delays from the average $\tilde{\tau}$ affects the phase of oscillations.

(ii) The repression/activation of transcription causes phase delay/lead of the following protein oscillations.

# 7 Conclusion

We have developed a systematic method to explore the profiles of oscillatory protein concentrations in biochemical networks with negative cyclic feedback. First, we have formulated harmonic balance for biochemical oscillators. Then, the relation between the reaction rates and the frequency, phase and amplitude of the oscillations has been analytically obtained. The proposed method has been demonstrated with the Pentilator and the somitogenesis clock, and the analysis result has shown quantitative agreement with the existing biological experiments.

A distinctive feature of the proposed approach is that the oscillation profiles are written in an analytic way. Thus, we can obtain detailed biological insight, which provides useful guidance in synthetic gene circuit design. We have revealed dimensionless parameters that primarily encode the profiles of oscillations, and analyzed the qualitative properties of frequency and phase in terms of these parameters.

**Acknowledgments:** This work was supported in part by the Ministry of Education, Culture, Sports, Science and Technology in Japan through Grant-in-Aid for Exploratory Research No. 21656106, Grant-in-Aid for Scientific Research (A) No. 21246067, and Grant-in-Aid for JSPS Fellows No. 23-9203.

# A Oscillation frequency for the case of heterogeneous parameters

In this section, we first discuss the oscillation frequency under parametric perturbations to $a_i, b_i, c_i$ and $\beta_i$ $(i = 1, 2, \cdots, N)$, then we show that Theorem 1, which is derived under Assumption 2, is useful for analyzing the upper and lower bounds of the frequency. To this end, we hereafter relax Assumption 2 and assume that the parameters belong to the following set $\mathcal{P}$.

$$\mathcal{P} := \{(a_i, b_i, c_i, \beta_i) \ (i = 1, 2, \cdots, N) \mid \underline{a_i} \leq a_i \leq \overline{a_i}, \underline{b_i} \leq b_i \leq \overline{b_i}, \\ \underline{c_i} \leq c_i \leq \overline{c_i}, \underline{\beta_i} \leq \beta_i \leq \overline{\beta_i} \ (i = 1, 2, \cdots, N)\}, \quad (46)$$

where the symbols with a upper and a lower bar are the given upper and lower bounds of each parameter, respectively.

We can verify that the frequency of oscillations can be obtained by (15) for the heterogeneous parameters as well. In particular, the following proposition shows that the upper and lower bounds of the frequency is obtained by the analysis for exte­reme parameters.

**Proposition 2.** *Consider the gene regulatory networks modeled by (1). Suppose the parameters $(a_i, b_i, c_i, \beta_i)$ $(i = 1, 2, \cdots, N)$ belong to $\mathcal{P}$. Then, the solution of (15) is bounded by $\underline{\varpi} \leq \varpi \leq \overline{\varpi}$, where $\underline{\varpi}$ and $\overline{\varpi}$ are the solutions of (15) for $(a_i, b_i) = (\underline{a_i}, \underline{b_i})$ and $(a_i, b_i) = (\overline{a_i}, \overline{b_i})$, respectively. The parameters $c_i$ and $\beta_i$ do not affect the solution of (15).*

**Proof.** We see that $\varpi$ satisfies

$$|I - H(j\varpi)K_1| = 0 \iff 1 + \prod_{i=1}^{N} h_i(j\varpi)|\xi_i| = 0. \quad (47)$$

This implies

$$\sum_{i=1}^{N} \angle h_i(j\varpi) = -\pi \iff \sum_{i=1}^{N} \left(-\varpi(\tau_{r_i} + \tau_{p_{i-1}}) - \angle(a_i + j\varpi) - \angle(b_i + j\varpi)\right) = -\pi. \quad (48)$$

We see from (48) that $\varpi$ monotonically increases as $a_i$ and $b_i$ increase and that $c_i$ and $\beta_i$ do not affect $\varpi$. These observations immediately lead to Proposition 2. □

Although this proposition shows that the parameters $c_i$ and $\beta_i$ do not affect the solution of (15), this does not mean the waveform of oscillations does not depend on these parameters, since the harmonic balance method is based on the approximation (10). In general, $c_i$ and $\beta_i$ affect the amplitude of oscillations, and the oscillations tend to contain high frequency components as $c_i$ and $\beta_i$ increase, which results in the degradation of the estimation by Theorem 1. Nevertheless, the qualitative properties are preserved for wide range of parameters even for such cases as noted in Remark 6.

Let $\underline{a} := \min_i \underline{a_i}$, $\underline{b} := \min_i \underline{b_i}$, $\overline{a} := \max_i \overline{a_i}$ and $\overline{b} := \max_i \overline{b_i}$. Then, it follows from Proposition 2 that the upper and lower bounds of the frequency can be obtained by letting $(a, b) = (\overline{a}, \overline{b})$ and $(a, b) = (\underline{a}, \underline{b})$ in Theorem 1, respectively. This means that Theorem 1, which is derived under Assumption 2, also provides some knowledge for the case of heterogeneous parameters.

# B The proof of Lemma 2

The poles of $\mathcal{H}_1(s)$ are obtained by solving

$$|\phi(s)e^{s\tau}I - UK_1| = |\phi(s)e^{s\tau}I - \Lambda| = 0, \quad (49)$$

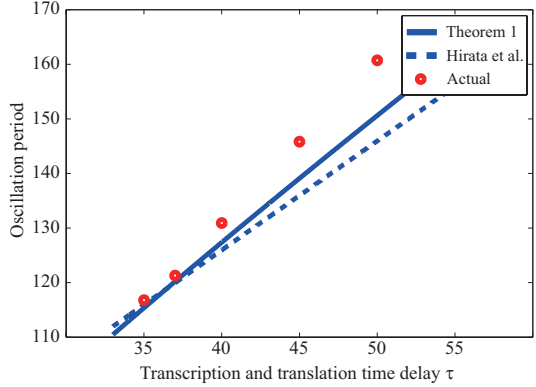

Figure 9: Comparison of the period of oscillations estimated from Theorem 1, Hirata et al. (2004) and the simulation of (40).

where $\Lambda := \mathrm{diag}(\lambda_1, \lambda_2, \cdots, \lambda_N) \in \mathbb{C}^{N \times N}$. It follows that

$$|\phi(s)e^{s\tau}I - \Lambda| = \prod_{i=1}^{N}(\phi(s)e^{s\tau} - \lambda_i)(=: \chi(s)). \tag{50}$$

Define $X(s) := \prod_{i=1, i \neq \ell, \ell'}^{N}(\phi(s)e^{s\tau} - \lambda_i)$. Then, we can write $\chi(s) = (\phi(s)e^{s\tau} - \lambda_\ell)(\phi(s)e^{s\tau} - \lambda_{\ell'})X(s)$. It should be noted that $\mathcal{H}_1(s)$ is a retarded time delay system, thus the dominant pole is located at the rightmost position in the complex plane.

In what follows, we prove the sufficient condition. It follows from the definition of $\Omega_+^c$ that the roots of $X(s) = 0$ lie in the open left half complex plane. On the other hand, $\phi(s)e^{s\tau} - \lambda_\ell = 0$ and $\phi(s)e^{s\tau} - \lambda_{\ell'} = 0$ have at least one root at $s = j\omega_0$ and $s = -j\omega_0$, respectively, and the rest in the open left half plane. In particular, the roots $s = \pm j\omega_0$ of these equations are single roots, because $\omega_0$ satisfies $|\phi(j\omega_0)e^{j\omega_0\tau}| = |\phi(j\omega_0)| = |\lambda_\ell|$ and $|\phi(j\omega)|$ monotonically increases with respect to $\omega > 0$. Therefore, $\mathcal{H}_1(s)$ has a pair of poles at $s = \pm j\omega_0$ with multiplicity one, and the rest in the open left half plane.

The proof of the necessary condition is omitted, since it can be easily obtained from the above discussion. □

## C Comparison of Theorem 1 and the estimate in Hirata et al. (2004)

In Hirata et al. (2004), the period of oscillations is analytically estimated as $2(\tau_r + \tau_p + \ln(2)/a + \ln(2)/b)$, where $a$ and $b$ are defined by (41). We here compare the period obtained from Theorem 1, Hirata et al. (2004) and numerical simulation of (40). The result is shown in Fig. 9.

We see that Theorem 1 and Hirata et al. (2004) provide close estimation for a wide range of time delay. The actual oscillation period starts to deviate from these estimations as the time delay increases, because the waveform tends to be distorted. Nevertheless, the relative error at $\tau = 50$ [min] is $-6.28\%$ for Theorem 1 and $-9.18\%$ for the Hirata's estimate. Thus, these estimates can be useful for gaining qualitative insight of the oscillation period. Note that Theorem 1 is useful for biochemical networks consisting of a large number of genes, while the estimate in Hirata et al. (2004) is applicable only to the self-negative feedback case, i.e. $N = 1$.

# D   Approximation of (44)

The right-hand side of (44) is rewritten as

$$\frac{1}{Q^2}\left\{\cot\left(\frac{\pi}{N}-\tilde{\varpi}\tilde{\tau}\right)\left(\sqrt{1+Q^2\tan^2\left(\frac{\pi}{N}-\tilde{\varpi}\tilde{\tau}\right)}-1\right)\right\}.$$

As $\tilde{\tau}\to\infty$, $\pi/N-\tilde{\varpi}\tilde{\tau}\to 0$ follows, thus $Q^2\tan^2(\pi/N-\tilde{\varpi}\tilde{\tau})$ becomes sufficiently small so that we can approximate the square root by $\sqrt{1+x^2}\simeq 1+x^2/2$. Then, we have

$$\tilde{\varpi}\simeq\frac{1}{2}\tan\left(\frac{\pi}{N}-\tilde{\varpi}\tilde{\tau}\right). \tag{51}$$

When $\pi/N-\tilde{\varpi}\tilde{\tau}$ is close to zero, $\tan(\pi/N-\tilde{\varpi}\tilde{\tau})\simeq\pi/N-\tilde{\varpi}\tilde{\tau}$. Thus, $\varpi\simeq 1/2(\pi/N-\varpi\tau)$, and the normalized period of oscillations is obtained as

$$\frac{2\pi}{\tilde{\varpi}}\simeq 4N+2N\tilde{\tau}=4N+2\sum_{i=1}^{N}(\tau_{r_i}+\tau_{p_i})\frac{1}{T_A}. \tag{52}$$